\documentclass[draftclsnofoot,12pt,onecolumn]{IEEEtran}
\newif\iftwocolumn\twocolumnfalse

\IEEEoverridecommandlockouts

\usepackage{amsmath}
\usepackage{dsfont}
\usepackage{setspace}
\usepackage{siunitx}
\DeclareSIUnit{\Bit}{Bit}
\DeclareSIUnit{\bit}{bit}
\DeclareSIUnit{\Bits}{Bits}
\DeclareSIUnit{\bits}{bits}
\usepackage{soul}
\usepackage{algorithm}
\usepackage{algpseudocode}
\usepackage{amssymb}
\usepackage{dblfloatfix}    % To enable figures at the bottom of page
\usepackage[normalem]{ulem} % sout
\usepackage{xcolor}
\usepackage[noadjust]{cite}
\usepackage{graphicx}
\usepackage{tikz}
\usepackage{lipsum}
\usepackage{bm}
\usepackage{pgfplots}
\usepackage{xfrac}
\ifCLASSOPTIONcompsoc
    \usepackage[caption=false, font=normalsize, labelfont=sf, textfont=sf]{subfig}
\else
\usepackage[caption=false, font=footnotesize]{subfig}
\fi

\usepackage{verbatim}
\pgfplotsset{compat=1.16}
\usetikzlibrary{plotmarks}
\usetikzlibrary{patterns}
\usetikzlibrary{spy}
\makeatletter
\def\centerarc[#1](#2)(#3:#4:#5)% Syntax: [draw options] (center) (initial angle:final angle:radius)
    { \draw[#1] ($(#2)+({#5*cos(#3)},{#5*sin(#3)})$) arc (#3:#4:#5); }

\DeclareMathOperator*{\argmax}{\mathrm{arg\,max}}

\usepackage{array}
\newcolumntype{L}[1]{>{\raggedright\let\newline\\\arraybackslash\hspace{0pt}}m{#1}}
\newcolumntype{C}[1]{>{\centering\let\newline\\\arraybackslash\hspace{0pt}}m{#1}}
\newcolumntype{R}[1]{>{\raggedleft\let\newline\\\arraybackslash\hspace{0pt}}m{#1}}
\newcommand{\customlabel}[2]{%
\protected@write \@auxout {}{\string \newlabel {#1}{{#2}{}}}}

\makeatother

\usepackage[normalem]{ulem}

\begin{document}
\title{Localization Attack by Precoder Feedback Overhearing in 5G Networks and Countermeasures}

\author{Stefan Roth$^1$, Stefano Tomasin$^2$,  Marco Maso$^3$, and Aydin Sezgin$^1$\\ 
\small{ $^1$Digital Communication Systems, Ruhr University Bochum, Germany,}\\
\small{$^2$Department of Information Engineering, University of Padova, Italy, $^3$Nokia Bell Labs, France.} \\
\small{\{stefan.roth-k21, aydin.sezgin\}@rub.de, stefano.tomasin@unipd.it, marco.maso@nokia.com
}
\thanks{This work has been funded in part by the Deutsche Forschungsgemeinschaft (DFG, German Research Foundation) under Germany's Excellence Strategy - EXC 2092 CASA - 390781972, and in part by MIUR (Italian Minister for Education) under the initiative {\it Departments of Excellence} (Law 232/2016).}
}
 \maketitle
 
\begin{abstract}
In fifth-generation (5G) cellular networks, users feed back to the base station the index of the precoder (from a codebook) to be used for downlink transmission. The precoder is strongly related to the user channel and in turn to the user position within the cell. We propose a method by which an external attacker determines the user position by passively overhearing this unencrypted layer-2 feedback signal. The attacker first builds a map of fed back precoder indices in the cell. Then, by overhearing the precoder index fed back by the victim user, the attacker finds its position on the map. We focus on the type-I single-panel codebook, which today is the only mandatory solution in the 3GPP standard. We analyze the attack and assess the obtained localization accuracy against various parameters. We analyze the localization error of a simplified precoder feedback model and describe its asymptotic localization precision. We also propose a mitigation against our attack, wherein the user randomly selects the precoder among those providing the highest rate. Simulations confirm that the attack can achieve a high localization accuracy, which is  significantly reduced when the mitigation solution is adopted, at the cost of a negligible rate degradation.
\end{abstract}


\section{Introduction}\label{sec:intro}

Economic and political interests push for the deployment of technologies to localize users, to the benefit of companies and states. As a matter of fact, many new services exploit the knowledge of our position, with a significant added value. For this reason, in the definition of the 5th generation (5G) of cellular systems, a significant emphasis has been placed on localization services \cite{LMS-rel15}. In order to localize a user, the cellular network may receive the radio signals transmitted by the user through one or more base stations, possibly equipped with multiple antennas \cite{8240645}. In general, for a higher transmission frequency, the localization error obtained with this technique is reduced, \cite{6779222}. Therefore, moving from microwaves in  4th generation (4G) networks (GHz) to millimeter waves (mmWave, at hundreds of GHz) in 5G networks, and then to sub-millimeter waves (THz) in  6th generation (6G) networks, paves the way for a localization precision within a few centimeters. 

Unfortunately, malicious devices external to the network may also be interested in localizing specific users, with a severe violation of their  privacy.  This is a key issue nowadays, with various personal and societal impacts \cite{doi:10.1080/17489725.2011.642820}, and  motivates the investigation carried out in this paper. Indeed, position privacy was indicated as one of the critical security issues, in particular of the new 5G networks \cite{8792139}.

One solution to localize the user is to exploit again its transmitted radio signal, similar to what is legitimately done by the cellular network. In this respect, various  localization methods are available in the literature based on the channel state information. From the channel state information, the time of arrival (ToA) or angle of arrival (AoA) can be estimated, and then trilateration or triangulation can be applied, see \cite{Wang2015} and references therein. Another approach  is also called radio fingerprint pattern matching  (RFPM) attack \cite{6244790,8891416}. The received signal strength indicator (RSSI)  \cite{7783292, 7218669}, and the multipath channel impulse response \cite{9234644} can also be used to this end. For a survey, see \cite{8409950} and \cite{WEN201921} for applications to massive MIMO in 5G networks. Among mitigation techniques against these attacks, we mention the privacy-preserving beamforming \cite{Tomasin-comml20}. 

In an alternative approach, the attacker eavesdrops on the demodulated signal coming from the user and the legitimate base stations. This type of attack may be easier to implement than that based on channel estimation, as it only requires demodulation and decoding  of digital signals. In particular, it does not require multiple receive antennas (typically needed for localization based on channel parameters) and can use standard chip-sets. Its lower complexity potentially makes it more dangerous for the  privacy of user location. In \cite{rupprecht-19-layer-two}, some attacks operating at layer 2 of the long-term evolution (LTE) protocol are presented, including  tracking of user positions and mapping of user temporary network identity to the temporary radio identity. Eavesdropping the LTE temporary identifiers enables tracking users also as they move from one cell to another \cite{Hong-18}.~\footnote{This attack is not possible anymore in 5G networks, where the temporary identifier is refreshed whenever the user changes  cell.}  In \cite{Shaik} and \cite{10.1145/3317549.3323416}, various attacks are proposed to retrieve the user position, by either letting the user report its global navigation satellite system (GNSS) position or obtaining the identifier of its serving base station. 

Focusing on truly passive solutions based on eavesdropping of control signals (not exploiting the reporting of the GNSS position), we note that all attacks in the literature provide only the information on the cell where the user is, which is a coarse localization. These attacks are also denoted as cell identifiers (CID) attacks. In this paper, we propose a novel localization attack, based on eavesdropping the control signal transmitted by users and related to their channel state information, as provided by the new radio (NR) 3GPP standard specifications of 5G networks. In particular, 3GPP provides means for the user to estimate the downlink channel and then choose the best precoder from a predefined codebook; lastly, the user feeds back to the base station the index of the selected precoder. Indeed, the precoder is related to the channel experienced by the user, and thus it reveals, in part, its position.  

We consider the following attack. The attacker (possibly supported by colluded users) first builds a map of precoder indices fed back from various positions in the cell. Then, by overhearing the precoder index fed back by the victim user, the attacker finds its position on the map. We focus on the  type-I single-panel codebook, which is the only mandatory solution in the standard, as of today. When operating at mmWaves, both the user and the base station must be equipped with multiple antennas, and a precise beamforming is needed for data communications, thus a more precise description of the precoder will be fed back. This, in turn, provides a better localization of the user, similar to what happens in radio localization techniques.

We analyze the attack and assess the obtained localization accuracy against various parameters, e.g., the feedback mode, the number of subbands, and the size of the cell. We analyze the asymptotic localization error of a simplified precoder feedback model as the number of subbands and channel clusters get large. We also propose a mitigation strategy against our attack, wherein the user randomly selects the precoder among those providing the higher rate. Extensive simulations on reference channel models confirm that the attack can achieve a high localization accuracy, which is instead significantly reduced when the mitigation solution is adopted, at the cost of a negligible rate degradation. Moreover, we compare our technique with the three aforementioned solutions available in the literature: the CID, the ToA, and the RFPM attacks. Our solution has a significantly better localization accuracy than the CID attack. Its accuracy improves as the number of subbands grows, outperforming ToA in most cases, while also requiring simpler hardware and software. Conversely, its accuracy is lower than that of RFPM, which however requires more complex hardware and software.

The rest of the paper is organized as follows. Section~\ref{sec:sysmodel} explains the system model of the channel and its spatial correlation. Section~\ref{sec:channelInformationFeedback} describes the precoder index feedback protocol defined by the 3GPP standard. The proposed localization attack is described in Section~\ref{sec:UELocalizationAttack}. The  mean square error of the localization attack is analyzed in Section~\ref{sec:performance}. The numerical results based on the standard channel model are given in Section~\ref{sec:NumericalResults}. Finally, Section~\ref{sec:Conclusion} concludes the paper.
 
\paragraph*{Notation}
In this paper, $\otimes$ denotes the Kronecker product. $\mathbb{E}[\cdot]$ denotes the expectation operator. Scalar variables are indicated in italics, while column vectors and matrices are in boldface italics, and sets are in calligraphic. $\lfloor x \rfloor$ and $\prec x \succ$ denote the largest integer $i \leq x$ and the nearest integer rounding $x$, respectively; $||\cdot||_2$ is the $L_2$-norm. ${\rm Q}(\cdot)$ denotes the Q-function.  $x \sim \mathcal U([a, b])$ means that the random variable $x$ is uniformly distributed  in the interval $[a,b]$; $\bm{x}\sim\mathcal{CN}(\bm{\mu},\bm{\Sigma})$ represents a circulant-symmetric Gaussian-distributed variable with mean $\bm{\mu}$ and variance $\bm{\Sigma}$. $\bm{I}_{N}$ is the identity matrix of size $N\times N$, whereas ${\rm det}[\bm{X}]$, $\bm{X}^T$, and $\bm{X}^H$ denote the determinant, transpose, and conjugate-transpose of matrix $\bm{X}$, respectively.  $|\mathcal X|$ denotes the cardinality of set $\mathcal X$. $\mathrm{atan2}(y,x)$ is the two-argument arctangent.

\section{Channel Model}\label{sec:sysmodel}
 
\begin{figure} 
    \centering
    \includegraphics{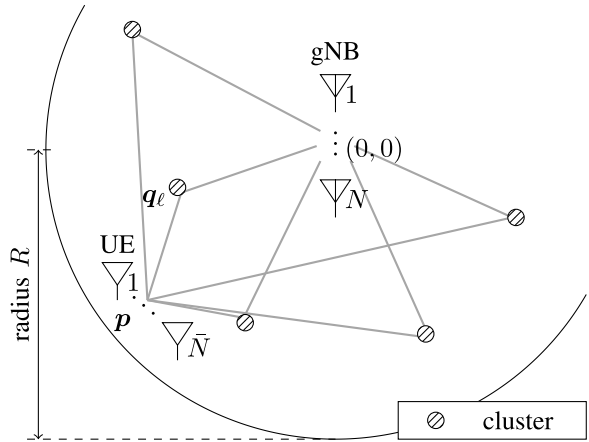}
    \caption{Schematic description of the channel model for a circular cell of radius $R$, and a generic cluster $\ell$, of coordinates $\bm{q}_{\ell}$. } \label{fig:schematic}
\end{figure}

As shown in \figurename~\ref{fig:schematic}, we consider a circular cell with radius $R$ of a cellular network.~\footnote{We consider here a two-dimensional space, ignoring its altitude, for the sake of a simpler explanation.} The new-generation node base (gNB) is located at the center of the circle, which coincides with the origin of coordinates (0,0), while the position of the {\em victim} user equipment (UE) has coordinates  $\bm{p}=(p_x,p_y)$. 

The gNB is equipped with a linear array of $N$ cross-polarized antennas ($2N$ antennas in total), whereas the UE is equipped with a linear array of $\bar{N}$ antennas. According to release 16 of the 3GPP cellular network standard, orthogonal frequency division modulation (OFDM) is used in downlink and subcarriers are grouped in subbands.

We assume that $K$ subbands are assigned to the victim UE $\bm{p}$ and that the channel is the same for all subcarriers within the same subband. Let $\bm{H}(\bm{p},k)$ be the downlink channel on subband $k$ for a UE in position $\bm{p}$. The gNB provides one stream (or layer in 3GPP parlance) to the UE at mmWave: the gNB transmits a digitally modulated symbol $s_{\kappa}(k)$ on subcarrier $\kappa$ of  subband $k$, with a precoding $N$-size column vector $\bm{w}(k)$, so that the  signal received at the UE is 
\begin{equation}
    \bm{r}_{\kappa}(k) = \bm{H}(\bm{p},k) \bm{w}(k) s_{\kappa}(k) + \bm{v}_{\kappa}(k), \quad k=1, \ldots, K,
    \label{iochannel}
\end{equation}
where $\bm{v}_{\kappa}(k)$ is the additive white Gaussian noise (AWGN) $\bar{N}$-size column vector with zero-mean independent entries and variance $\sigma_v^2$.~\footnote{To improve the readability, we use letters $v$ and $V$ consistently for noise variables, whereas  letter $w$ is for precoding vectors.}

We now introduce the channel model. First, note that we do not adopt the  model defined by the 3GPP standard \cite{3GPP:15}, since therein the channel realizations do not depend on the UE position. In reality though, the angle of departure (from the gNB) and of arrival (at the UE) of each channel path are related to the position of the gNB, the UE, and the reflectors that define the path. As we exploit this information to localize the UE, we need a  spatially-consistent and location-consistent channel model. Therefore, we first recall the location-consistent model of \cite{1033686} and then propose a modification to make it  also spatially-consistent. Note that this model is not tailored to the proposed localization method, but it only  realistically connects the wireless channel characteristics to the position of the UE and surrounding objects. Such a model is therefore useful for the analysis of any localization method based on the channel features.

\subsection{Location-consistent Channel Model}

We now provide the channel model for each subband, recalling that all subcarriers of the same subband experience the same channel.  According to \cite{1033686}, the channel between a UE and its serving gNB is determined by reflections on $L$ {\em clusters}, assumed to be static. The position of cluster $\ell \in \mathcal{L} = \{1, \ldots, L\}$ has coordinates $\bm{q}_{\ell}=(q_{x,\ell},q_{y,\ell})$.  This scenario is illustrated in \figurename~\ref{fig:schematic}.

The transmission delay of the path through cluster $\ell$ is 
\begin{align}
    \tau_{\ell}(\bm{p})&=\frac{1}{c_0}\Big(\left|\left|\bm{p}-\bm{q}_{\ell}\right|\right|_2+\left|\left|\bm{q}_{\ell}\right|\right|_2\Big),\label{eq:distancesToTau}
\end{align}
where $c_0$ is the speed of light. Then, according to \cite{8240645}, the channel complex gain for all subcarriers of subband $k\in\mathcal{K}=\{1,\dots,K\}$, relative to cluster $\ell$,  is
\begin{equation}
    \gamma_{\ell}(\bm{p},k)=\sqrt{N}\frac{A_{\rm PL} g_{\ell}(k)}{\tau_{\ell}(\bm{p})}e^{-j\frac{2\pi k\tau_{\ell}(\bm{p})}{KT_{\mathrm{S}}}}, 
    \label{beta}
\end{equation}
where $A_{\rm PL}$ is the path-loss normalization factor,  $T_{\mathrm{S}}$ is the sampling period, and  $g_{\ell}(k)$ is zero-mean unitary-variance complex Gaussian distributed and independent for each subband and each cluster. Gain $g_{\ell}(k)$ is also assumed to be time-invariant.

The angle between cluster $\ell$ and the gNB is 
\begin{equation}
\phi'(\bm{q}_{\ell})=\mathrm{atan2}\left(q_{y,{\ell}},q_{x,{\ell}}\right),
\end{equation}
while the angle between cluster $\ell$ and the UE is 
\begin{equation}
\bar{\phi}'(\bm{p},\bm{q}_{\ell})=\mathrm{atan2}\left(q_{y,\ell}-p_y,q_{x,\ell}-p_x\right).
\end{equation}
For subband $k$, the angle of departure from the gNB is \cite{3GPP:15}
\begin{equation}
    \phi(\bm{p},\bm{q}_{\ell},k)=\phi'(\bm{q}_{\ell})+c_{\mathrm{ASD}}\alpha_{\ell}(\bm{p},k),\label{eq:aod_los}
\end{equation}
while the angle of arrival to the UE is
\begin{equation}
    \bar{\phi}(\bm{p},\bm{q}_{\ell},k)=\bar{\phi}'(\bm{p},\bm{q}_{\ell})+c_{\mathrm{ASA}}\bar{\alpha}_{\ell}(\bm{p},k)+v_{\mathrm{O}},
\label{eq:aoa_los}
\end{equation}
where $\alpha_{\ell}(\bm{p},k) \sim \mathcal U([-1, 1])$ and $\bar{\alpha}_{\ell}(\bm{p},k)  \sim \mathcal U([-1, 1])$ are spatially correlated (see Section~\ref{sec:spatialcorr}) and independent per subband, and $c_{\mathrm{ASD}}$ and $c_{\mathrm{ASA}}$ are constants \cite[Table 7.5-3]{901}; lastly, $v_{\mathrm{O}} \sim \mathcal U([0, 2\pi))$ represents the orientation of the device in the $x$-$y$-plane.

For a signal departing from the gNB into the direction of cluster $\ell$ at subband $k$, $k=1, \ldots, K$, let us define the steering vectors, representing the relative phase shift among antennas as \cite{8240645} 
\begin{align}
    \bm{a}_{\ell}(\bm{p},k)&=\frac{1}{\sqrt{N}}\begin{pmatrix}e^{j\frac{2\pi d}{\lambda_k}\left(-\frac{N-1}{2}\right)\sin \phi(\bm{p},\bm{q}_{\ell},k) }\\\vdots\\e^{j\frac{2\pi d}{\lambda_k}\left(\frac{N-1}{2}\right)\sin \phi(\bm{p},\bm{q}_{\ell},k) }\end{pmatrix},
    \label{eqai}
\end{align} 
where $d$~\footnote{Throughout the paper, all distances are denoted with the letter $d$.} is the antenna spacing at the gNB, 
\begin{equation}
\lambda_k = \frac{c_0}{f_{\mathrm{c}} + \Delta_f (k - (K-1)/2)},
\end{equation}
is the wavelength at subband $k$, $f_{\mathrm{c}}$ is the central carrier frequency, and $\Delta_f$ is the subband spacing.
Similarly, for a signal received at the UE, the steering vectors are
\begin{align}
    \bm{\bar{a}}_{\ell}(\bm{p},k)&=\frac{1}{\sqrt{\bar{N}}}\begin{pmatrix}e^{j\frac{2\pi \bar{d}}{\lambda_k}\left(-\frac{\bar{N}-1}{2}\right)\sin \bar{\phi}(\bm{p},\bm{q}_{\ell},k) }\\\vdots\\e^{j\frac{2\pi \bar{d}}{\lambda_k}\left(\frac{\bar{N}-1}{2}\right)\sin \bar{\phi}(\bm{p},\bm{q}_{\ell},k) }\end{pmatrix},
\end{align}
where $\bar{d}$ is the UE antenna spacing. Lastly, the narrowband channel at subband $k$ is modeled as the $\bar{N}\times 2N$ matrix \cite{8240645}  
\begin{align}
    \bm{H}(\bm{p},k)=\sum_{\ell \in\mathcal{L}}\gamma_{\ell}(\bm{p},k) \bm{\bar{a}}_{\ell}(\bm{p},k)\left(\begin{pmatrix}\cos \mu \\\sin \mu\end{pmatrix}\otimes\bm{a}_{\ell}(\bm{p},k)\right)^H\label{eq:channelMatrix},
\end{align}
where $\mu \sim \mathcal U([0, 2\pi))$ is the  co-phasing angle, related to the orientation of the UE with respect to the gNB. Note that (\ref{eq:channelMatrix}) is the stacked version of two channel vectors of length $N$, each corresponding to one of the two polarizations of the $2N$ antennas. 

This model is location-consistent, since any UE in position $\bm{p}$ will experience reflections from (fixed) clusters, thus having  the same $\gamma_{\ell}(\bm{p},k)$ and $\bm{a}_{\ell}(\bm{p},k)$. Only the co-phasing angle $\mu$ is independent for each UE in the same position, as the orientation of each UE does not depend on its position.

\subsection{Channel Spatial Correlation}
\label{sec:spatialcorr}
Channels of UEs at different positions are spatially correlated, as confirmed by a vast literature,~\cite{7999256, 8647188,8690738,8570034, 901}. Therefore, all UE channels are determined by the same clusters while other parameters are changing slowly in space. In order to obtain a  spatially consistent model, we assume  $\alpha_{\ell}(\bm{p},k)$ to be spatially correlated at different locations $\bm{p}$, thus (see \cite{901})
\begin{equation}
    \mathbb E[\alpha_{\ell}(\bm{p}_1,k)\alpha_{\ell}(\bm{p}_2,k)] =  e^{-\frac{||\bm{p}_1 - \bm{p}_2||_2}{2d_{\rm S}}},
    \label{unifcorr}
\end{equation}
where $d_{\rm S}$ is the spatial correlation distance. Note that $\{\alpha_{\ell}(\bm{p},k)\}$ are  spatially correlated (in $\bm{p}$), while still being independent in $\ell$ and $k$. 

\section{Precoder Information Feedback}
\label{sec:channelInformationFeedback}

The precoder $\bm{w}(k)$ to be used on subband $k$ (see \eqref{iochannel}) is determined by the UE, according to its channel conditions. To this end, the gNB periodically transmits a pilot signal to the UE. From the received signal,  the UE obtains the {\em channel estimate} at subband $k$, modeled as  
\begin{align}
    \hat{\bm{H}}(\bm{p},k)=\bm{H}(\bm{p},k)+\bm{V},\label{eq:channelMatrixNoise}
\end{align}
where matrix $\bm{V}$ represents the AWGN estimation error vector, with  independent zero-mean complex Gaussian distributed entries, i.e.,  $(\bm{V})_{\bar{n},n}\sim\mathcal{CN}(0,\sigma^2)$, $\bar{n}=1, \ldots, \bar{N}$, $n = 1, \ldots, 2N$. 

Based on $\hat{\bm{H}}(\bm{p},k)$, the UE selects the precoder $\bm{w}(k)$ from a codebook shared with gNB, as detailed in the following. Note that  3GPP  \cite{rpt:3gpp.38.214} does not mandate a precise algorithm to select the precoding vectors. Here, we assume that the  UE aims at maximizing the  downlink throughput achieved with the selected precoder. Assuming perfect modulation and coding schemes, the spectral efficiency of each subcarrier of subband $k$ can be estimated from (\ref{iochannel}) as
\begin{equation}
\begin{split}
    {\mathcal R}& \left[\{\bm{w}(k)\}\right] =\\
    & \sum_{k=1}^K\log_2{\rm det}\left[\bm{I}_{\bar{N}}+\frac{\hat{\bm{H}}(\bm{p},k)\bm{w}(k)\bm{w}(k)^H\hat{\bm{H}}(\bm{p},k)^H}{\sigma_v^2}\right].
    \end{split}
    \label{ratest}
\end{equation}
Note that (\ref{ratest}) is an {\em estimate} of the spectral efficiency, while the actual spectral efficiency has a similar expression, with $\hat{\bm{H}}(\bm{p},k)$ replaced by $\bm{H}(\bm{p},k)$.  

Once the UE has selected the precoding vectors $\{\bm{w}(k)\}$, it transmits to the gNB a sequence of bits, identifying them within the codebook.  The 3GPP standard \cite{rpt:3gpp.38.306}  defines a large codebook and then specifies two procedures for the selection of vector and the feedback of their indices. We denote these two procedures as {\em feedback modes}.

In the rest of this section, first, we describe the codebook construction, and then the two feedback modes. Lastly, we describe a new feedback mode, not included in the standard, that will be useful in the performance analysis of our localization technique.

\subsection{Codebook Construction}

Several codebook types are defined in release 16 of the 3GPP standard \cite{rpt:3gpp.38.214}. In first NR deployments, UEs  will be mandated to support only the type-I single-panel codebook, while other codebook types are optional \cite{rpt:3gpp.38.306}. Therefore, here we focus on this codebook. 

Each precoding vector is identified by the couple $(m,n)$, where $m \in \mathcal{M} = \{0, 1, \ldots, NO-1\}$, wherein $O$ is denoted as {\em oversampling factor},  and $n \in \mathcal{N} = \{0,\dots,3\}$. Index $m$ identifies the vector  
\begin{align}
    \bm{\Tilde{w}}_m&=\left(1,\ \exp\left(\frac{j2\pi m}{NO}\right),\ \dots,\ \exp\left(\frac{j2\pi (N-1)m}{NO}\right)\right)^T.
    \label{eqv}
\end{align}
Index $n$ is associated with the co-phasing angle between the polarization of antennas couples $\psi_n=\exp\left(j2\pi n/4\right)$. Then, the precoding vector associated with the couple $(m,n)$ is defined as~\footnote{Each codebook vector has also a horizontal and vertical component. As the antenna array is linear in the considered scenario, the vertical component is fixed to $1$. Therefore, only the horizontal component is present in each codebook vector.} 
\begin{align}
    \bm{w}_{m,n}&=\frac{1}{\sqrt{N}}\begin{pmatrix}\bm{\Tilde{w}}_m\\\psi_n\bm{\Tilde{w}}_m\end{pmatrix}&m\in\mathcal{M},\ n\in\mathcal{N}.
    \label{defw}
\end{align}
Note that the construction of the tuples $(m,n)$ varies through the different feedback modes.

\subsection{Feedback Mode 1}

In feedback mode $1$, the same precoding vector is used for all subbands, whereas a subband-dependent co-phasing of the two polarizations is applied. The UE selects the precoder as the solution of the following rate maximization problem:
\begin{equation}
    \begin{split}
    &(m^\star,\{n^\star(k)\}) =\argmax_{\substack{m,\{n(k)\}}}  {\mathcal R}\left[\{\bm{w}(k) = \bm{w}_{m,n(k)}\} \right],
    \label{eq:rate}
    \end{split}
\end{equation}
where  $m\in\mathcal M$ and $n(k)\in\mathcal{N}$, for each subband $k$.

The feedback to the gNB comprises two indices, $i_{1,1}$ and $i_2$, which provide subband-independent regarding subband-dependent information, respectively. For feedback mode 1, they are defined as follows:
\begin{itemize}
    \item $i_{1,1}$ is the $\log_2(NO)$-bits unsigned integer representation of  $m^\star \in \mathcal{M}$;
    \item $i_2(k)$, $k=1, \ldots, K$, is the 2-bits unsigned integer representation of  $n^\star(k)$. This requires the feedback of $2K$ bits overall. 
\end{itemize}

\subsection{Feedback Mode 2}

In feedback mode $2$, the choice of both precoding vectors and co-phasing of the two polarizations is subband-dependent.~\footnote{We are considering here the subband mode of operation of \cite{rpt:3gpp.38.214}.}
The co-phasing of the two polarizations is chosen as for feedback mode $1$. Then, the UE selects the precoder by solving the throughput maximization problem 
\begin{equation}
\begin{split}
    &(m^\star, \{\delta^\star(k)\},\{n^\star(k)\})  =  \\
    &\hspace{1.5cm} \argmax_{m, \{\delta(k)\},\{n(k)\}} {\mathcal R}\left[\{\bm{w}(k) = \bm{w}_{2m + \delta(k),n(k)}\} \right], 
\end{split}\label{eq:rate2}
\end{equation}
where  $m\in\{0,\dots,(NO)/2-1\}$ and  $\delta(k)\in\{0,\dots,3\}$ for each subband.

About the feedback information we have:
\begin{itemize}
    \item $i_{1,1}$ is a $(\log_2(NO)-1)$-bits    unsigned integer carrying the index $m^\star$;
    \item $i_2(k)$, $k=1, \ldots, K$, is a 4-bits packet, whose  2 most and 2 least significant bits carry the (unsigned integer) value of  $\delta^\star(k)$ and   $n^\star(k)$, respectively. This requires the feedback of $4K$ bits overall. 
\end{itemize}

\subsection{Feedback Mode 3}
 
In order to dig deeper into the comparison of position errors for a different number of subbands, we consider a non-standard feedback procedure, where we report a different precoding vector for each subband, thus with different values of both $m(k)$ and $n(k)$.  We  find the couple of indices for each subband $k=1, \ldots, K$ as 
\begin{equation}
    \begin{split}
    &(\{m^\star(k)\}, \{n^\star(k)\}) = \\ &\hspace{2cm}\argmax_{\substack{\{m(k)\},\{n(k)\}}}
      {\mathcal R}\left[\{\bm{w}(k) = \bm{w}_{m(k),n(k)} \right].
    \label{eq:rate3}
    \end{split}
\end{equation}

About the feedback information we have:
\begin{itemize}
    \item $i_2(k)$, $k=1, \ldots, K$, is an unsigned integer, whose the $\log_2(NO)$ most significant bits represent $m^\star(k)$ and the 2 least significant bits represent of $n^\star(k)$. This requires the feedback of $K(\log_2(NO)+2)$ bits overall. 
\end{itemize}

Note that,  for $K=1$, all considered feedback modes coincide. 

\section{UE Localization Attack and Mitigation} \label{sec:UELocalizationAttack}

We now first characterize the attacker, and then describe how the attack is performed, in order to localize the victim UE. Lastly, we propose a mitigation strategy against the attack.

\subsection{Attacker Model}

The attacker is a device able to overhear the control plane signal exchange between the gNB and the UE. In particular, here we assume that the attacker eavesdrops the precoder feedback of the UE, i.e., $i_{1,1}$ and $\{i_2(k)\}$. 

As observed in Section II, UEs in the same position experience different  polarizations and noise, yielding different values of $i_{1,1}$ and $\{i_2(k)\}$. As the attacker is interested only in the UE position, he will neglect the information related to  polarization, namely indices $n$ of $\bm{w}_{m,n}$ (see \eqref{defw}). This corresponds to completely ignoring $i_2(k)$ in feedback mode~1, and the two least significant bits of $i_2(k)$ in feedback modes~2 and 3.

In the following,  $\bm{b}(\bm{p}) = [b_1, \ldots, b_K]$~\footnote{For readability, we denote all variables related to feedback bits with the letters $b$, $B$, and $\beta$.}  denotes the vector of  bits useful for localization (obtained from $m^{\star}(k)$ or $m^\star$ with or without $\delta^\star(k)$) fed back in position $\bm{p}$ for each subband $k$,  $k=1, \ldots, K$. In addition, $\mathcal B$ is the set of possible values of $b_k$.  Note that  the quantization process is still affected by noise, yielding different feedback vectors $\bm{b}(\bm{p})$ from the same location $\bm{p}$.

\subsection{Proposed Localization Attack}

We propose a novel UE localization attack that comprises two {\em stages}: 1) the {\em preliminary stage}, wherein the attacker creates a map that associates to each position $\bm{s}$ in the cell the bit vector $\bm{b}(\bm{s})$ fed back from there, and 2) the {\em attack stage}, wherein the victim UE is localized by matching its fed back bit vector $\bm{b}$ to the map constructed in the preliminary stage. We now provide a detailed description of the two stages.

\paragraph{Preliminary Stage} We assume that, in the preliminary stage, the attacker  benefits from the collaboration of colluded UEs that move within the cell and collect the precoder feedback bits. Further observations on this assumption are provided in Section \ref{observationsec}. Let $\mathcal S$ be the set of explored positions in this stage. Depending on the effects of noise, even for the same position,  we can collect different feedback bit vectors. Let us define the probability of occurrence of vector $\bm{b}(\bm{s}) = \bm{\beta}$, $\bm{s} \in \mathcal S$, as
\begin{equation}
     {\tt p}_{\rm map}(\bm{\beta},\bm{s}) = \mathds{P}[\bm{b}(\bm{s}) = \bm{\beta}]\,. 
\end{equation}
This probability provides a probabilistic map of feedback vectors to positions in the cell and will be exploited by the attacker to localize the victim UE in the next stage. We also define the probability of occurrence of feedback vector $\bm{\beta}$ as
\begin{equation}
    {\tt p}(\bm{\beta}) = \mathds E[\,\mathds{P}[\bm{b}(\bm{p}) = \bm{\beta}]\,]\,, 
\end{equation}
where the expectation is taken with respect to the position $\bm{p}$. 

\paragraph{Attack Stage} In the attack stage, the attacker overhears the fed back bit vector $\bm{b}^*$ sent by the victim UE from an unknown position. Then, the attacker finds the position of the UE by averaging all points having as fed back bits $\bm{b}^*$, i.e., 
\begin{equation}
    \hat{\bm{p}}(\bm{b}^*) = (\hat{p}_x, \hat{p}_y) =   \sum_{\bm{s}\in\mathcal{S}} \, \frac{{\tt p}_{\rm map}(\bm{b}^*,\bm{s})}{{\tt p}(\bm{b}^*)}\bm{s}. \label{recpos}
\end{equation}

We observe that a suitable estimate of the probabilities $ {\tt p}_{\rm map}(\bm{\beta},\bm{s})$ and ${\tt p}(\bm{\beta})$ can be obtained from the collected samples in the preliminary stage. 

\subsection{Potential Practical Implementation of the Proposed Attack}\label{observationsec}
Some comments are needed now on the proposed attack.
First, we observe that the proposed attack is different from the localization solutions based on a trilateration of channel parameters, such as angles of arrival and departure and RSSI. Indeed, negligible signal processing is needed or performed by the attacker on the fed back bit vector, since only a simple mapping is required. To this end, the attacker may be equipped with a single antenna only, whereas multiple antennas are needed for an accurate localization by trilateration. Moreover, off-the-shelf chip-sets for the decoding of control signals can be used for the proposed attack without dedicated hardware; the software part is also very elementary, being only a database of positions and feedback vectors.

In terms of the computational complexity entailed with the construction of the map, we observe that, when the feedback bits are collected by colluded UEs in a known position, minor data processing, is required. Still, it may require some time and the availability of colluded users to extensively explore the cell. The localization precision is also dictated by the density of points of the map for which a correspondence between the position and the feedback bits is available. Moreover, the map should be refreshed whenever the wireless propagation environment changes. As an alternative, the attacker may infect neighboring UEs (but not the victim UE itself to exclude the trivial case) with a virus application that sends the current UE position to the attacker, which then associates it to the feedback bits~\cite{Shaik}.

Another approach may include the use of machine-learning techniques, such as self-organizing maps \cite{kohonen1982self}, for automatic construction of the map, even without knowledge of user position. Note also that the correspondence between positions and feedback bits changes over time due to the movement of objects, therefore the map should be periodically updated. The two solutions (virus injection and use of self-organizing maps) can be run continuously, providing a constant update of the map.

\subsection{Mitigation Solutions}

In order to mitigate the effects of the attack based on feedback eavesdropping, one possibility is to use  an asymmetric encryption algorithm when transmitting the feedback vector $\bm{b}(\bm{p})$. With this solution, only a ciphertext can be eavesdropped, such that different locations are indistinguishable for the attacker. However, this solution yields an encryption overhead and requires a change of the standardized transmission protocol.

We consider a second mitigation option, that can be implemented with the current standard. The UE first selects the $U$ precoders in the codebook providing the highest rates. Then, the UE chooses uniformly at random one among the $U$ precoders, and feeds back its index to the gNB. Therefore, multiple  precoding vectors (and fed back bit vectors) will be associated to the same position, 
decreasing the localization accuracy of the attack. Consider for example the feedback mode~1, and let $m^\star_u$, $u=1, \ldots, U$, be distinct feedback values (remember that for localization purposes $n^\star$ is not useful), i.e.,
\begin{equation}
   m^\star_{u_1}  \neq m^\star_{u_2}  \quad \mbox{for } u_1 \neq u_2,\; u_1, u_2 \in \{1, 2, \ldots, U\}
\end{equation}
for which the rates on subband $k$ are in decreasing order, i.e., 
\begin{equation}
\begin{split}
    {\mathcal R}[\{\bm{w}(k) &= \bm{w}_{m^\star_{1},n_{1}^\star(k)}\}] \geq  \ldots\geq {\mathcal R}[\{\bm{w}(k) = \bm{w}_{m^\star_{U},n_{U}^\star(k)}\}]\,.
\end{split}
\end{equation}
Now, the UE generates $u$ with equal probability in the set of integers $1, \ldots, U$, and feeds back $(m^\star_u,\{n^\star(k)\}_u)$. Clearly, on one hand the localization will be less accurate. On the other hand, the UE incurs into a rate reduction of forthcoming data downlink transmissions, due to the suboptimal choice of the precoding vector.

\section{MSE Performance Analysis}
\label{sec:performance}

The effectiveness of our attack can be assessed in terms of the mean square error (MSE) of the position estimation, i.e.,
\begin{equation}
    \mathrm{MSE} = {\mathbb E}[||\hat{\bm{p}}- \bm{p}||^2],\label{eq:MSENumerical}
\end{equation}
or equivalently, in terms of the root MSE $\mathrm{RMSE} = \sqrt{\mathrm{MSE}}$.

We now derive the MSE of the position estimation in a specific scenario. We have the following assumptions on the UE and the gNB:
\begin{enumerate}
     \item\label{enum:assumptionCbm3} use of feedback mode 3;
    \item\label{enum:assumptionEstim} perfect channel estimation at the UE;
    \item\label{enum:assumptionPolar} fixed UE polarization with $\mu =\pi/4$. 
\end{enumerate}
Assumption \ref{enum:assumptionCbm3}) is useful for understanding the impact of the feedback for multiple subbands. By assumptions \ref{enum:assumptionEstim}) and \ref{enum:assumptionPolar}), there will be a deterministic feedback vector $\bm{b}(\bm{p})$  from each location $\bm{p}$.  

About the channel model, we further assume:
\begin{enumerate}
\setcounter{enumi}{3}
    \item\label{enum:assumptionAlpha} random and independent $\alpha_{\ell}(\bm{p},k)$ at each position ($d_{\mathrm{S}}=0$);
    \item\label{enum:averaassumption} random and independent $\alpha_{\ell}(\bm{p},k)$ at each channel observation time; 
    \item\label{enum:assumptionRecei} one receive antenna ($\bar{N} = 1$);
    % \item\label{enum:assumptionGain} {\color{blue} \sout{random and independent channel gains $g_{\ell}(k)$ for each realization (even when UE  position is unchanged);}}
    \item\label{enum:assumptionOneCl} channel on subband $k$, $k=1, \ldots, K$, determined only by the cluster $\ell^\star(k)$, having the highest channel gain;
    \item\label{enum:assumptionuniform} uniform distribution of UEs in the cell area.  
\end{enumerate}
Assumption \ref{enum:assumptionAlpha})  removes the spatial consistency of $\alpha_{\ell}(\bm{p},k)$. With assumption \ref{enum:averaassumption}) we assume that $\alpha_{\ell}(\bm{p},k)$ changes even when the UE stands (or pass again) in the same position. Therefore, as the attacker collects samples to build the probabilistic map  ${\tt p}_{\rm map}(\bm{\beta}, \bm{s})$, the value of  $\alpha_{\ell}(\bm{p},k)$ changes. Under  assumption \ref{enum:assumptionRecei}), the selected beamformer is the (quantized version of the) maximal ratio transmit beamformer, which, under assumptions \ref{enum:assumptionPolar}) and \ref{enum:assumptionOneCl}) is 
\begin{equation}
\bm{w}(k) = \left[\begin{matrix}
\bm{a}_{\ell^{\star}(k)}(\bm{p},k) \\
\bm{a}_{\ell^{\star}(k)}(\bm{p},k) 
\end{matrix}\right].     
\end{equation}
Assumptions \ref{enum:assumptionRecei}) and \ref{enum:assumptionOneCl}) are strictly related: indeed, when using multiple antennas at the receiver, two clusters may be exploited. Hence, the beamformer chosen according to (\ref{eq:rate2}) will not be matched to either of the two, but it will split the signal power between the two, also according to the clusters' gains. 
Lastly, about assumption \ref{enum:assumptionOneCl}), from  (\ref{beta}) and (\ref{eq:channelMatrix}), we can compute the index of the cluster having the highest channel gain. In particular, indicating the   distance between the gNB and cluster  $\ell$ as $d_{\ell}  = \left|\left|\bm{q}_{\ell}\right|\right|_2$, and   the  distance between position $\bm{p}$ and cluster  $\ell$ as $\bar{d}_{\ell}(\bm{p}) =  \left|\left|\bm{p}-\bm{q}_{\ell}\right|\right|_2$, let us define
\begin{equation}
    \zeta_\ell(\bm{p},k) =  \frac{\xi_{\ell}(k)}{\left(\bar{d}_{\ell}(\bm{p})+d_{\ell}\right)^2},
    \label{defzeta}
\end{equation}
and 
\begin{equation}
    \xi_{\ell}(k) = |g_{\ell}(k)|^2. 
\end{equation}
Then, the index of the cluster chosen from subband $k$ is obtained at the maximum of $\gamma_{\ell}(\bm{p},k)$ with respect to $\ell$, i.e., from (\ref{beta}) and ignoring irrelevant constants,
\begin{equation}
    \ell^{\star}(k) = \argmax_{\ell \in\mathcal{L}} \zeta_\ell(\bm{p},k).
    \label{istar}
\end{equation}
In this case, we say that the UE is {\em linked} to cluster $\ell^{\star}(k)$.
 
\subsection{Conditional MSE for Given Cluster Positions and Gains}

First, we derive the feedback statistics for given cluster positions $\{\bm{q}_{\ell}\}$ and gains $\{\xi_{\ell}\}$. For given cluster positions, the distance between the gNB and cluster $d_{\ell}$ is deterministic, and, for a given UE position $\bm{p}$, also the distance between cluster and the UE, $\bar{d}_{\ell}(\bm{p})$ is deterministic. 

Let $\bm{X}(\bm{p},\{\bm{q}_{\ell}\},\{\xi_{\ell}\},k)$ be a $NO$-size vector, whose entry $X_{i}(\bm{p},\{\bm{q}_{\ell}\},\{\xi_{\ell}\},k)$ is the probability that the UE on subband $k$ (linked to cluster $\ell^{\star}(k)$) feeds back $i_2(k) = i$. We now consider Assumptions \ref{enum:assumptionAlpha}) and \ref{enum:averaassumption}), and include the effects of $\alpha(\bm{p},k)$ in our analysis. In particular, since the selected beamformer is the (quantized version of the) maximal ratio combining beamformer to $\bm{a}_{\ell^{\star}(k)}(\bm{p},k)$, from \eqref{eq:aod_los}, \eqref{eqai}, and \eqref{eqv}, we have 
\begin{equation}
\begin{split}
    &X_{i} (\bm{p},\{\bm{q}_{\ell}\},\{\xi_{\ell}\},k) \\&= 
    {\mathbb P}\left[\frac{d}{\lambda_k} \sin  \phi(\bm{p},\bm{q}_{\ell^\star(k)},k)   \in \left[\frac{i-0.5}{NO}-\bar{o}, \frac{i+0.5}{NO}-\bar{o}\right)  \right],% \\
    %= &  \mbox{\hl{TBD}}.
    \end{split}\label{eq:Qli}
\end{equation}
where $\bar{o}\in\{0,1\}$.
For the computation of (\ref{eq:Qli}), let us define
\begin{align}
    \theta_1(i,k,\tau,\bar{o}) &=  \frac{1}{c_{\mathrm{ASD}}} \bigg\{\arcsin\left[\frac{\lambda_k(i-0.5-\bar{o}NO)}{dNO}\right]_{-1}^{1}\nonumber\\&\hspace{2.1cm} -(-1)^{\tau}\phi'\left(\bm{q}_{\ell^\star(k)}\right)-\tau\pi \bigg\},
\\
    \theta_2(i,k,\tau,\bar{o}) &= \frac{1}{c_{\mathrm{ASD}}} \bigg\{\arcsin\left[\frac{\lambda_k(i+0.5-\bar{o}NO)}{dNO}\right]_{-1}^{1}\nonumber\\&\hspace{2.1cm} - (-1)^{\tau}\phi'\left(\bm{q}_{\ell^\star(k)}\right)-\tau\pi\bigg\},
\end{align}
where, $[x]_{a_1}^{a_2}=\min(\max(x,a_1),a_2)$. From \eqref{eq:aod_los} and recalling that  $\alpha_{\ell^{\star}(k)}(\bm{p},k) \sim \mathcal U([-1, 1])$, after some algebraic steps, we obtain the close-form expression
\begin{align}
    &X_{i}(\bm{p},\{\bm{q}_{\ell}\}, \{\xi_{\ell}\},k) \nonumber\\&\hspace{1.1cm}= \frac{1}{2}\sum_{\tau\in\mathbb{Z}}\sum_{\bar{o}=0}^1\left|\left[[\theta_1(i,k,\tau)]_{-1}^1, [\theta_2(i,k,\tau)]_{-1}^1\right)\right|.
    \label{eq:Qli2}
\end{align}

For a feedback bit-vector $\bm{b}$,   let also 
\begin{equation}
B(\bm{b}) = \sum_{k=1}^K (NO)^{k-1} b_k.
\end{equation}
Since the channel gains are independent per subband,  $b_k$ will be independent for $k=1, \ldots, K$, for given positions. Hence, the column vector $\bm{Y}(\bm{p},\{\bm{q}_{\ell}\})$ of length  $(NO)^K$ of the joint probability mass function (PMF) of the  feedback bit vector has entries 
\begin{align}
    Y_{B(\bm{b})}(\bm{p},\{\bm{q}_{\ell}\},\{\xi_{\ell}\})=\prod_{k=1}^{K} X_{i_2(k)}(\bm{p},\{\bm{q}_{\ell}\}, \{\xi_{\ell}\},k), \; \bm{b} \in \mathcal B.
    \label{Ycbref}
\end{align}
The average joint PMF over all UE positions is
\begin{align}
    \bar{Y}_{B(\bm{b})}(\{\bm{q}_{\ell}\},\{\xi_{\ell}\})=\frac{1}{A} \int  Y_{B(\bm{b})}(\bm{p},\{\bm{q}_{\ell}\},\{\xi_{\ell}\}) d\bm{p},
\end{align}
where $A = \int d\bm{p}=\pi R^2$ is the cell area and $\bm{b} \in \mathcal B$.

The reconstruction MSE for given cluster positions and gains is obtained as 
\begin{align}
    &\mathrm{MSE}(\{\bm{q}_{\ell}\},\{\xi_{\ell}\}) \nonumber\\&\hspace{0.5cm}=  \frac{1}{A}\sum_{\bm{b}}   \int ||\bm{p} - \hat{\bm{p}}(\bm{b})||^2 Y_{B(\bm{b})}(\bm{p}, \{\bm{q}_{\ell}\},\{\xi_{\ell}\}) d\bm{p}\,.
    \label{resMSE}
\end{align}
This integral can be computed by numerical methods.

\subsection{MSE Expectation}

Assuming that, in the preliminary stage, the UE collects the feedback bits from all positions, (\ref{recpos}) becomes   
\begin{equation}
      \hat{\bm{p}}(\bm{b}) =  \mathbb E[\bm{p}] = \int \bm{p} \frac{Y_{B(\bm{b})}(\bm{p}, \{\bm{q}_{\ell}\},\{\xi_{\ell}\})}{A\bar{Y}_{B(\bm{b})}(\{\bm{q}_{\ell}\},\{\xi_{\ell}\})} d\bm{p},
\end{equation}
where division by $A$ comes from the uniform distribution of UE position $\bm{p}$ in the cell.

The average MSE with respect to cluster and gNB positions is obtained by integrating (\ref{resMSE}) over the PDF of cluster positions and gains, i.e., 
\begin{align}
    \mathrm{MSE} &= \iint \cdots \iint \frac{\mathrm{MSE}(\{\bm{q}_{\ell}\}, \{\xi_{\ell}\})}{A^L} \left(\prod_{\ell=1}^{L}\prod_{k=1}^{K} \exp(\xi_l(k))\right)\nonumber\\&\hspace{4cm} 
    d \bm{q}_1d \xi_1  \cdots d \bm{q}_Ld \xi_L.\label{eq:MSELower}
\end{align}
If the clusters are assumed to be within the cell, the integrals are over the entire cell area and for a circular cell. The integrals can be computed by numerical methods.

\subsection{Asymptotic Analysis}
\label{sec:asymptotics}

Since the explicit computation of the MSE and thus the RMSE in \eqref{eq:MSELower} requires the solution of integrals by numerical methods, it is also interesting to explore the asymptotic behavior of the RMSE, as the number of subbands or clusters of gNB antennas goes to infinity. 

First, we consider the case wherein no feedback is available. Thus, the attacker has no specific information on the position of  the victim UE, apart from the knowledge that it is uniformly distributed in the cell. From the circular shape of the cell, the minimum MSE position is  the center of the cell.  Therefore, in this extreme case of $K=0$, the RMSE (denoted ${\rm RMSE}_0$) turns out to be 
\begin{align}
    \mathrm{RMSE}_{0}=\sqrt{\frac{1}{A}2\pi\int_{0}^{R}R^2RdR}=\sqrt{\frac{1}{A}\frac{2\pi}{4}R^4}=\frac{1}{\sqrt{2}}R.
    \label{MSE0}
\end{align}
Note that this value also describes the accuracy of a localization technique based on the user's temporary network identity when not involving tracking. With our approach, there will always be more feedback information available than what is considered in this approximation. Hence, \eqref{MSE0} is an upper-bound of the RMSE of our attack. 

Next, let us consider the case wherein the attacker overhears a  feedback from the victim UE, and in particular the feedback mode~3 is employed. We also consider a gNB equipped with a large number of antennas, i.e.,  $N \rightarrow \infty$. By approximating $\xi_\ell(k)$ with its mean $\xi_\ell(k) \approx \mathbb{E}[\xi_\ell(k)]$, from  \eqref{defzeta} and \eqref{istar}, we have that the UE is always linked to the cluster with the shortest path-length and each   cluster leads to a different feedback index. Under the assumption of a large number of clusters ($L \rightarrow \infty$) and a uniform distribution of the clusters in the circle, we can approximate the resulting regions assigned to the individual precoders as Voronoi regions around the clusters as circles of radius $r\approx R/\sqrt{L}$. Hence, when the number of gNB antennas goes to infinity, we localize the UE into the Voronoi region of its nearest cluster, \eqref{eq:MSELower}, thus the RMSE can be approximated as
\begin{equation}
    \mathrm{RMSE}_{\infty} = \lim_{L,N \rightarrow \infty} \mathrm{RMSE}(K) \approx \frac{R}{\sqrt{2L}}.
    \label{inftycluster}
\end{equation}

For intermediate values of $N>0$, we localize the UE by overlapping $K$ maps, each with $NO$ regions. Although maps are different for each subband, they all are obtained from the same cluster positions, being strongly related to each other. Thus, the RMSE decreases exponentially with decaying factor
\begin{align}
    \lim_{K \rightarrow \infty}     \frac{ \log\mathrm{RMSE}(K)} {K} = -\frac{1}{\eta} \log NO,
     \label{inftyK}
\end{align}
where $\eta$ takes into account the correlation of maps among subbands.

Lastly, from  \eqref{MSE0},  \eqref{inftycluster}, and \eqref{inftyK}, we have  
\begin{align}
    \mathrm{RMSE} &\approx \frac{R}{\sqrt{2L}}+\left(1-\frac{1}{\sqrt{L}}\right)\frac{R}{\sqrt{2}}\left(NO\right)^{-K/\eta}.
    \label{RMSEfit}
\end{align}
Note that this result is not based on the assumptions done at the beginning of this section. Moreover, this analysis can also be applied to feedback modes~1 and 2, by properly adjusting $\eta$. In addition to the correlation of maps at the various subbands, $\eta$ can also take into account the information feedback reduction due to correlation in the choice of precoders among subbands in modes 1 and 2. 
In the next section, we will assess the accuracy of this asymptotic result for various system configurations.

\section{Numerical Results}\label{sec:NumericalResults}

\begin{table} 
\centering
\caption{Parameter And Their Values}
\begin{tabular}{|p{1cm}|R{1.7cm}||p{1cm}|R{1.7cm}|}
\hline
$c_{\rm ASA}$&$\num{10}\degree$&$c_{\rm ASD}$&$\num{10}\degree$\\\hline
$f_c$ & $\SI{28}{G\Hz}$ & $\Delta_f$ & $\SI{5.76}{M\Hz}$ \\\hline
$K$&$4$&$L$&70\\\hline
$N$&$\{2,16\}$&$O$&$4$\\\hline
$P_x$&$100$&$P_y$&$100$\\\hline
$R$&$\SI{25}{\metre}$& $d_{\rm S}$ &$\SI{2}{\meter}$\\\hline
$L_{\mathrm{S}}$&$\SI{1}{\meter}$ & $d/\lambda$ & 0.5 \\\hline
$\sigma_w$& $1$ & & \\\hline\end{tabular}
\label{tab:numericalParameters}
\end{table}

We consider the scenario of Section \ref{sec:sysmodel}, with the parameters reported in Table~\ref{tab:numericalParameters}. Note that $K=13$ subbands correspond to a $\SI{74.88}{MHz}$ channel bandwidth, with 48 subcarriers per subband \cite{rpt:3gpp.38.214}.

The spatial correlation of $\alpha_{\ell}(\bm{p},k)$ has been obtained as follows. First,  we define a grid having square meshes on an area containing the cell, with points $\bm{g}(n,m) = (n \Delta_{\rm x}, m\Delta_{\rm y})$, $n \in \{-G_{\rm x}+1, \ldots, G_{\rm x}-1\}$, $m \in \{-G_{\rm y}+1, \ldots, G_{\rm y}-1\}$, $\Delta_{\rm x} = \eta R/G_{\rm x}$, $\Delta_{\rm y} = \eta R/G_{\rm y}$, and $\left|\left|\bm{g}(n,m)\right|\right|_2\leq \eta R$. For each point of the grid we generate independent Gaussian random variables $\{Z(n,m)\}$, with zero mean and unitary variance. Then, we filter  $\{Z(n,m)\}$ with the 2D filter   \cite{generationLi}
\begin{equation}
    Z(\bm{p}) = \sum_n \sum_m Z(n,m) \sin\left[\frac{\pi}{6} e^{-\frac{||\bm{p}-\bm{g}(n,m)||_2}{2d_{\rm S}}}\right].
\end{equation}
Lastly, in order to obtain a uniform distribution for $\alpha_{\ell}(\bm{p},k)$, we apply the transformation  \cite{generationLi}
\begin{equation}
    \alpha_{\ell}(\bm{p},k) = 2(1-{\rm Q}(Z(\bm{p})) - 1\,.
\end{equation}

By considering the {\em preliminary stage} of the attack,  the explored positions are inside the circle of radius $R$ on a square lattice $\bm{s} \in \mathcal S$, with
\begin{equation}
\begin{split}
    \mathcal S  &=  \big\{\bm{s}=(L_{\mathrm{S}} n_x, L_{\mathrm{S}} n_y)^T\, 
    \big|\,\left|\left|\bm{s}\right|\right|_2\leq R,n_x \in\mathbb{Z},n_y \in\mathbb{Z}\big\},
    \end{split}
\end{equation}
where $L_{\mathrm{S}}$ is the lattice spacing.

\begin{figure} 
\centering
\includegraphics{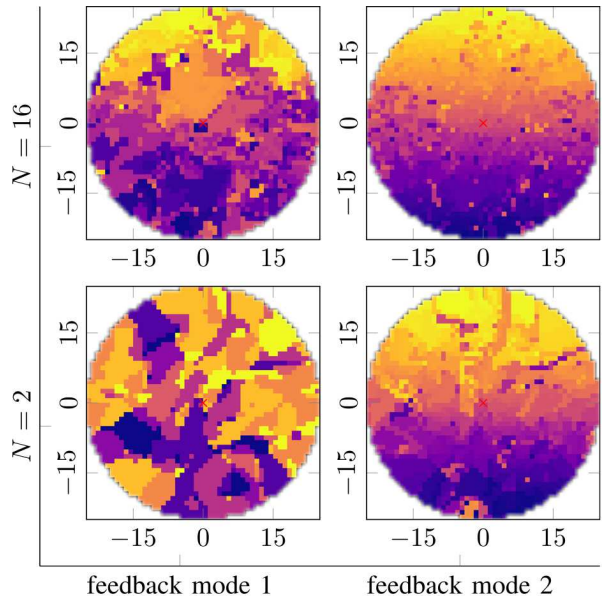}
\caption{Precoding vector index for each location in color-code, for different feedback modes and $N$, $K=4$ subbands, and without measurement noise.}
\label{fig:colormapPlots}
\end{figure}

As we mentioned, for each position $\bm{p}$ we have multiple associated feedback bit vectors, depending on the orientation (polarization) of the UE. Fig.~\ref{fig:colormapPlots} shows, by an associated color, the index of the most probable feedback bit vector $\bm{b}(\bm{p})$, collected in  the preliminary phase of the attack in each position $\bm{p}$, for an example instance of the simulated scenario. The four plots show maps for feedback modes~1 and 2, and $N=2$ and $16$, in the absence of measurement noise ($\sigma=0$). 

For  feedback mode~1 and $N=2$, the large monochrome regions indicate a low localization accuracy, as all points with the same color have the same feedback vector, thus are not distinguishable by the attacker. Also, note that some feedback vectors have a higher occurrence than others, which further decreases the accuracy, assuming a uniform UE distribution in space. Clearly, increasing the number of antennas $N$ helps, as the codebook gets larger:  for feedback mode~1 and $N=16$, we see more colors, while we still see various points with the same feedback vector. 

Feedback mode~2 further increases the variety of colors and yields a more homogeneous distribution of colors on the map, from which we expect a much better localization accuracy.  

\subsection{RMSE of the Disclosed Position}\label{sec:rmseReconPos}

\begin{figure} 
\centering
\includegraphics{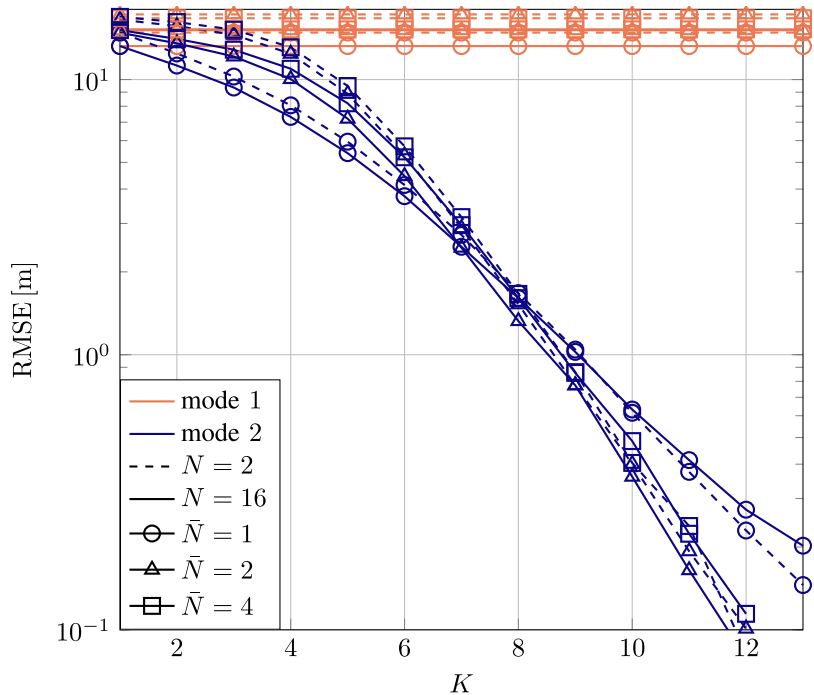}
\caption{RMSE vs the number of subbands $K$ for different feedback modes (mode) and numbers of transmit ($N$) and receive ($\bar{N}$) antennas, in the absence of measurement noise ($\sigma=0$).}
\label{fig:MseOversubbands}
\end{figure}

{\color{black}}We have simulated our attack with a map of lattice $\mathcal{S}$. Fig.~\ref{fig:MseOversubbands} shows the RMSE as a function of the number of subbands $K$ for different feedback modes and numbers of transmit antennas $N$.  This figure has been obtained in the absence of measurement noise ($\sigma=0$), while other parameters are as in Table~\ref{tab:numericalParameters}.

\begin{figure*}
\centering
\includegraphics{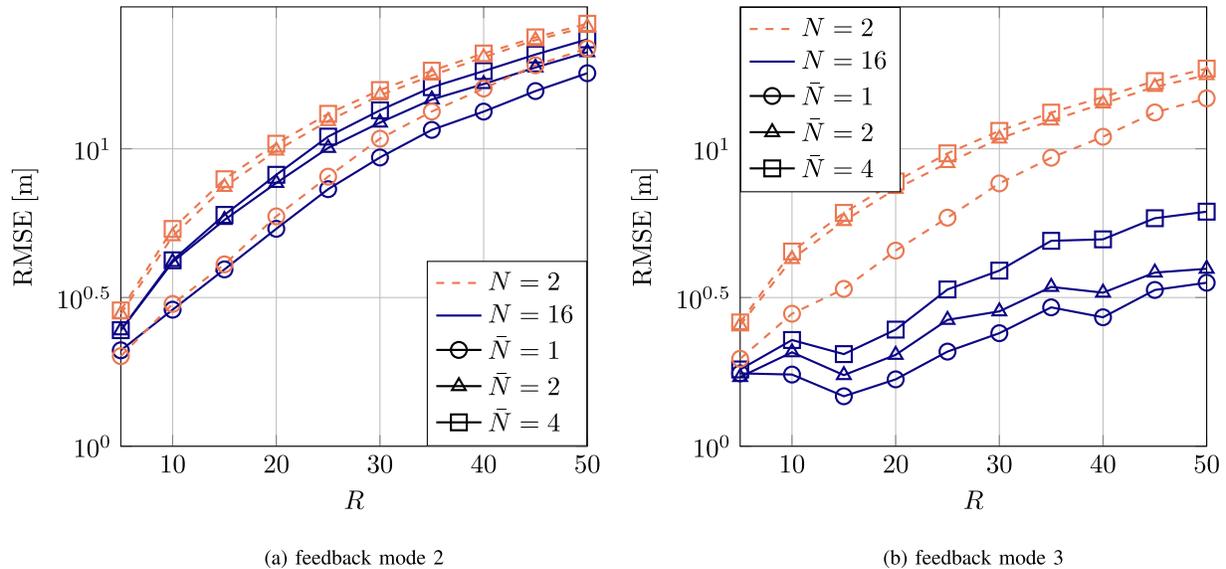}
%\vspace{-0.5cm}
\caption{RMSE over the cell-edge radius $R$, from left to right for feedback modes 2 and 3 for different numbers of transmit and receive antennas and $K=4$ subbands.}
\label{fig:mseOverArea}
\end{figure*}

Regarding feedback mode~1, as it only reports a single precoder for all subbands, the resulting position RMSE does not vary with $K$. Also, feedback mode~1 has a much higher RMSE than feedback mode~2, as also expected from the maps of Fig.~\ref{fig:colormapPlots}. Indeed, the RMSE of feedback mode~1 even slightly increases as we increase the number of antennas (thus increasing the number of feedback bits): although the codebook size increases, the additional bits are used for a better description of the single precoder for all subbands, thus not being useful for localization purposes.  We then conclude that using feedback mode~1 is actually an effective solution to obtain location privacy.

More interesting results are obtained with feedback mode~2. In this case,   using more subbands significantly lowers the RMSE, with an exponentially decaying behavior, as predicted by the analysis of Section~\ref{sec:asymptotics}. We also note that a slightly better localization is achieved  by increasing the number of antennas. In particular,  multiple-antenna UEs ($\bar{N}>1$) outperform devices with a  single antenna, for a large value of $K$. This can be explained as follows. From the resulting precoding vectors with $\bar{N}$ columns, the attacker can identify reflections on $\bar{N}$ clusters, thus having more accurate information on the victim location, implicitly doing a triangulation/trilateration as $\bar{N} >1$. Clearly, also by increasing $N$,  the RMSE (for a large $K$) decreases as the codebook gets larger and precoding vectors become denser, revealing more details on the channel, and thus on the UE location. However, note that, for small values of $K$, having more antennas at the receiver (larger $\bar{N}$) actually worsens the localization.  With feedback mode~2 and a large number of subbands $K$, we can achieve a location precision (in the absence of noise) significantly below $\SI{1}{\meter}$.

\figurename~\ref{fig:mseOverArea} shows the RMSE as a function of the cell radius $R$, for the different feedback modes and different numbers of antennas. Here, we let the average density of clusters per unit area unchanged with respect to the other figures, by setting the number of clusters  $L = \prec 0.112\, R^2\succ$, thus $L=70$ for $R=\SI{25}{\meter}$ as from Table~\ref{tab:numericalParameters}.  We observe that the RMSE increases with the radius as the cell map becomes larger while the size of the codebook does not increase. Indeed, as the feedback does not increase, the precision in associating feedback bits to positions is reduced for a larger cell. Regarding the different modes and number of antennas, the observations are similar to those obtained for the previous simulation scenarios.

\begin{figure} 
\centering
\includegraphics{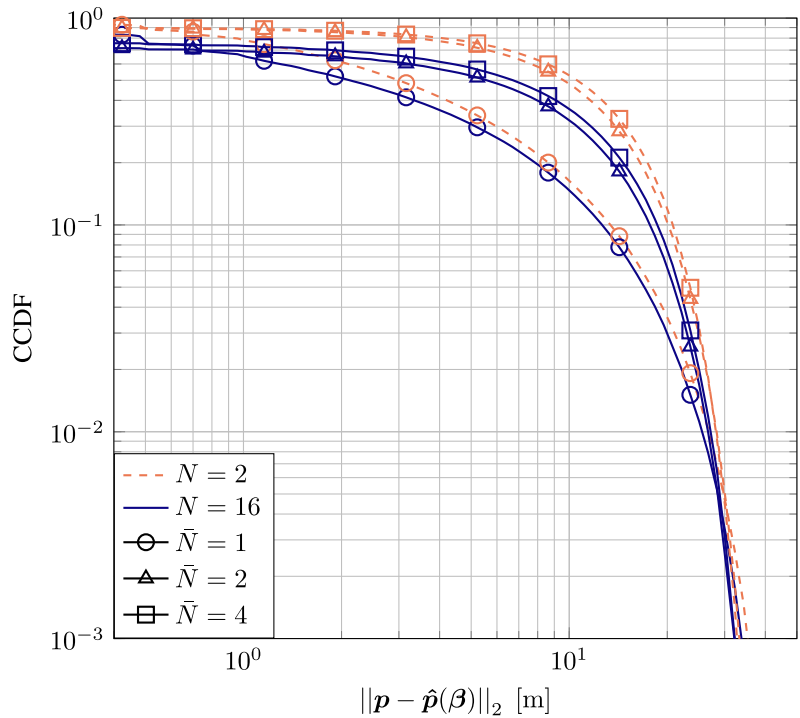}
\caption{CCDF of the error, for feedback mode 2 and different numbers of transmit antennas.}
\label{fig:msePdf}
\end{figure}

\subsection{Uncertainty of the Disclosed Position}

\figurename~\ref{fig:msePdf} shows the complementary cumulative density function (CCDF) of the estimation error in the absence of noise, for different feedback modes and numbers of antennas. The CCDF ${\mathbb P}[{\rm RMSE} > a]$ can also be read as the average percentage of points in the map with the same color (thus feedback bits), being at a distance larger than a threshold $a$ from the reconstructed position, i.e.,
\begin{equation}
\left|\left|\bm{p}-\hat{\bm{p}}(\bm{\beta})\right|\right|_2 > a,    
\end{equation}
where $\bm{b}(\bm{p}) =\bm{\beta}$. For $N=16$, $\bar{N}=1$, and $K=4$, in less than 20\% of the cases, we have a localization error of more than $\SI{10}{\meter}$, and in most cases, the localization error is significantly smaller.

\subsection{Comparison with Analytical Results}

\begin{figure}
    \centering
    \includegraphics{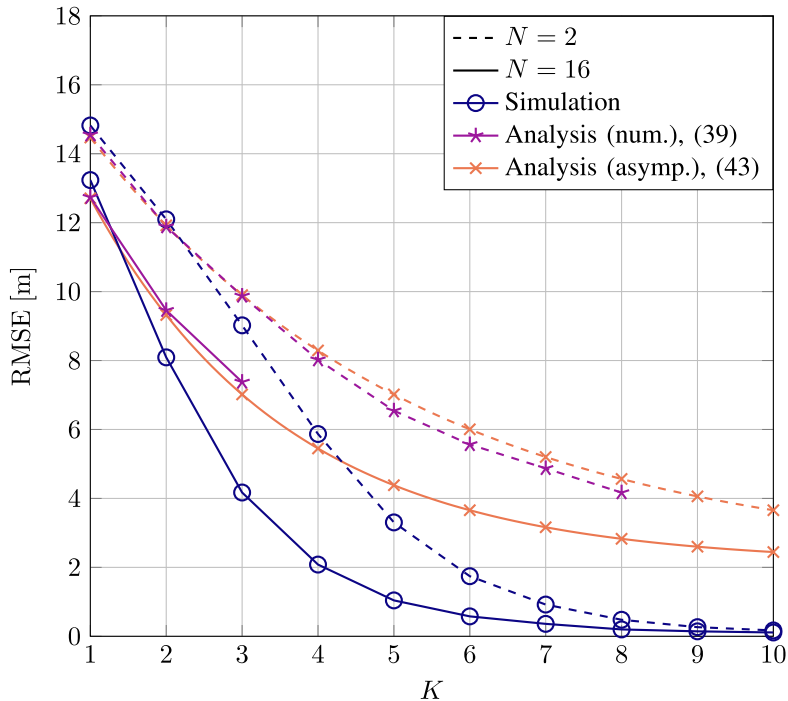}
    \caption{RMSE for feedback mode~3 for different numbers of transmit antennas $N$, vs the number of subbands,  with the general model of Sections~\ref{sec:sysmodel} and \ref{sec:channelInformationFeedback} (circle mark), and with the analytical results of Section~\ref{sec:performance} (star and cross marks).}
    \label{newfigure}
\end{figure}

We now focus on the analysis of Section~\ref{sec:performance}. Fig.~\ref{newfigure} shows the RMSE obtained with the general model of Sections~II and III (circle mark) and with the analytical results of Section~IV (star and cross marks). For the asymptotic results of \eqref{RMSEfit}, we set $\eta = 9$, which provides a good fit with the analytical results for both values of $N$. Note that we only show the RMSE \eqref{eq:MSELower} up to $K=3$ for $N=16$ since the computational complexity of \eqref{Ycbref} becomes excessive for higher values of $K$, as $B(\bm{b})$ takes $(NO)^K$ possible values.

Since the analytical model of Section~IV has various additional assumptions on both the channel and  the precoding feedback, we expect some difference to the simulation results obtained with the general model. The major difference  comes from the fact  that the analysis is obtained for a channel  determined by a single cluster $\ell^\star(k)$ in each subband (assumption \ref{enum:assumptionOneCl})). In the simulation results, instead, the optimal angle of departure is a combination of the angles of departure of multiple clusters. Thus, a larger number of angles can be optimal. Indeed, the RMSE obtained with \eqref{eq:MSELower} is an approximation of the RMSE obtained with the general model results.

The asymptotic approximation \eqref{RMSEfit}  shows a good match with the RMSE  \eqref{eq:MSELower} for both values of $N$. Also,  Fig.~\ref{newfigure}  validates that for low values of $K$, the asymptotic approximations are matching the numerical results well. For large values of $K$, the analytical results can be seen as an upper bound. This confirms its validity and practical usefulness.

\begin{figure*}
\centering
\includegraphics{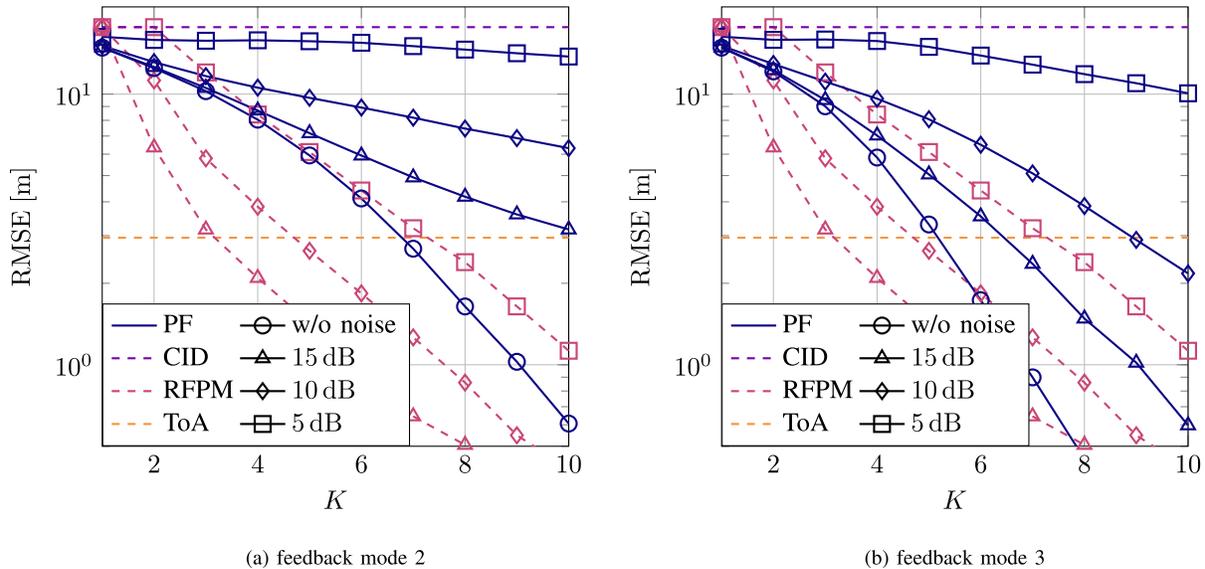}
%\vspace{-0.5cm}
\caption{RMSE over the number of subbands $K$ from left to right for feedback modes 2 and 3 for $N=2$ and $\bar{N} = 1$ while measurement noise is present. The $\si{\dB}$ values refer to the SNR.
}
\label{fig:MseOversubbandsNoise2}
\end{figure*}
\begin{figure*}
\centering
\includegraphics{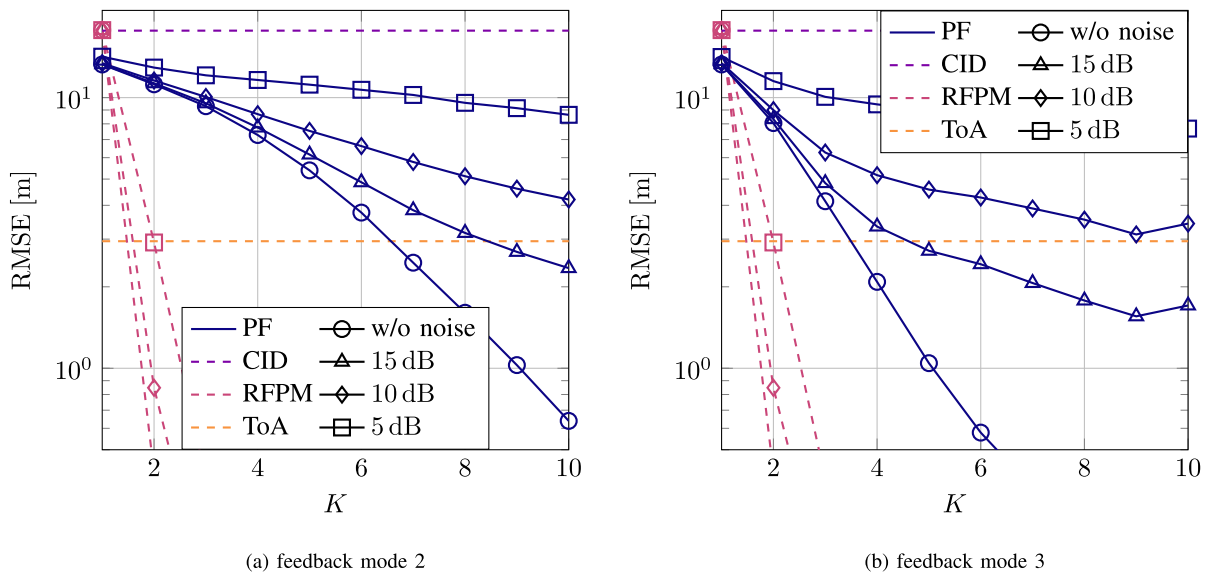}
%\vspace{-0.5cm}
\caption{RMSE over the number of subbands $K$ from left to right for feedback modes 2 and 3 for $N=16$ and $\bar{N} = 1$ while measurement noise is present. The $\si{\dB}$ values refer to the SNR.
}
\label{fig:MseOversubbandsNoise16}
\end{figure*}

\subsection{Comparison With Reference Localization Techniques}\label{sec:comparRefTechn}

We now compare the proposed localization attack with three reference solutions, namely a CID, a ToA and an RFPM attacks. Our proposed attack is denoted as precoder feedback (PF) attack.

In case of the CID attack, only the cell serving the victim user is identified, without any further location specification. This attack requires the eavesdropping of control signals that enable to identify the user, as in  \cite{rupprecht-19-layer-two} and \cite{Hong-18}. Moreover, it requires the knowledge of the gNB position and the cell area. Similarly to our PF attack, the CID attack operates on control signals. However, as already observed in Section V, the localization RMSE of CID, i.e., $\mathrm{RMSE}_0$ of \eqref{MSE0}, is rather large and indeed always larger than the localization RMSE of PF.

ToA attack utilizes three adversarial receivers, uniformly distributed within the cell area. Based on the ToA measurement at each receiver, the linear least squares method LLS-II-RS discussed in \cite{Wang2015} is used to localize the UE. Herein, we consider the ToA of the shortest path between UE and the receiver, through any of the clusters. In the simulated model, the locations of the adversarial receivers are distributed uniformly within the cell.~\footnote{We ensure that the three receivers are not located within one line by excluding all cases in which the difference between the angles of any receiver to the two others is smaller than $0.01\pi$.}

The RFPM attack \cite{6244790,8891416} is based on the estimation of the channel, which is subsequently compared to a database with cell positions to obtain the most likely estimate. We consider an attacker equipped with $2N$ antennas, which is identical to the legitimate gNB. To cope with the  random orientation of the UE the time-reversal resonating strength (TRRS) algorithm is used to compare the observed uplink channel estimate \cite{7029113,8891416} with those in the database.
This technique is based on overhearing reference signals transmitted by the UE, to estimate the uplink channel. This attack requires (as the PF attack) a preliminary stage, by means of which channels features are mapped to locations; for this reason, we included it in our comparison. However, differently from the PF attack, RFPM also requires sophisticated hardware and specialized signal processing capabilities and solutions for channel estimation, including the use of multiple antennas.

The RMSEs of the CID, ToA, RFPM, and proposed PF methods are compared in \figurename~\ref{fig:MseOversubbandsNoise2} and \figurename~\ref{fig:MseOversubbandsNoise16},  for different numbers of subbands $K$ and  both feedback modes of PF (cfr. Section~\ref{sec:channelInformationFeedback}). The RMSE of feedback mode 3 is illustrated for values of  signal-to-noise ratio (SNR) at the cell border, defined as follows
\begin{equation}
    {\rm SNR}   =  \frac{ {\mathbb E}_{i,j}\left|\bm{H}(\bm{p}_{R},k)\right|_{i,j}^2}{\sigma^2},
\end{equation}
where $\bm{p}_{R}$ is a position on the cell border. In particular, we consider ${\rm SNR}\in\{5, 10, 15\}\,\si{\dB}$ and noise-free reception (referred to as w/o noise in \figurename~\ref{fig:MseOversubbandsNoise2} and \figurename~\ref{fig:MseOversubbandsNoise16}). A degradation of the location accuracy as the SNR decreases is evident. Switching the focus to the accuracy delivered by the PF method for different feedback modes, we note that the PF attack with feedback mode~3 yields a lower RMSE than mode~2, thanks to the larger amount of information on each subband provided by the former.

When comparing the PF attack with CID, ToA, and RFM attacks, we note that the proposed attack significantly outperforms the CID attack, by reducing the RMSE from tens of meters to less than $\SI{1}{\meter}$, as the SNR and the number of subbands $K$ increase. The ToA attack achieves an accuracy of only a few meters, thus being outperformed by PF for large enough values of $K$. The lowest RMSE values are observed with the RFPM attack, which  provides a precision of about $\SI{1}{\meter}$, even when only $2$ subbands are used. This is expected, since RFPM  exploits the unquantized channel state information, differently from  PF, which operates on quantized feedback information. However, the better performance of RFPM comes at the cost of more complex hardware and software, including the need for more antennas.

\subsection{Attack Mitigation}
 
\begin{figure}
\centering
\includegraphics{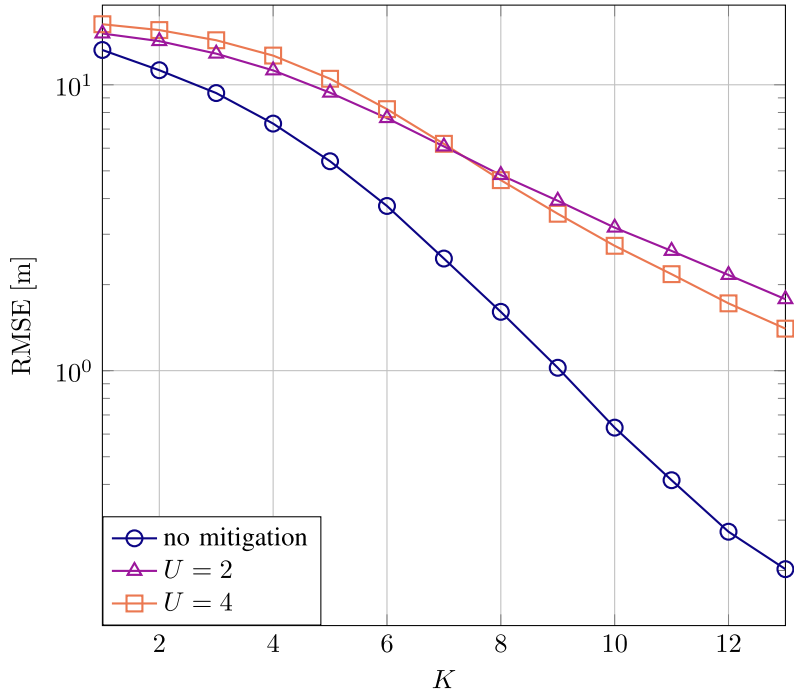}
\caption{RMSE vs $K$ with attack mitigation, for $U=2$ and 4, and comparison with the RMSE achieved without the attack mitigation.}
\label{fig:mitigationRmse}
\end{figure}

\begin{figure}
\centering
\includegraphics{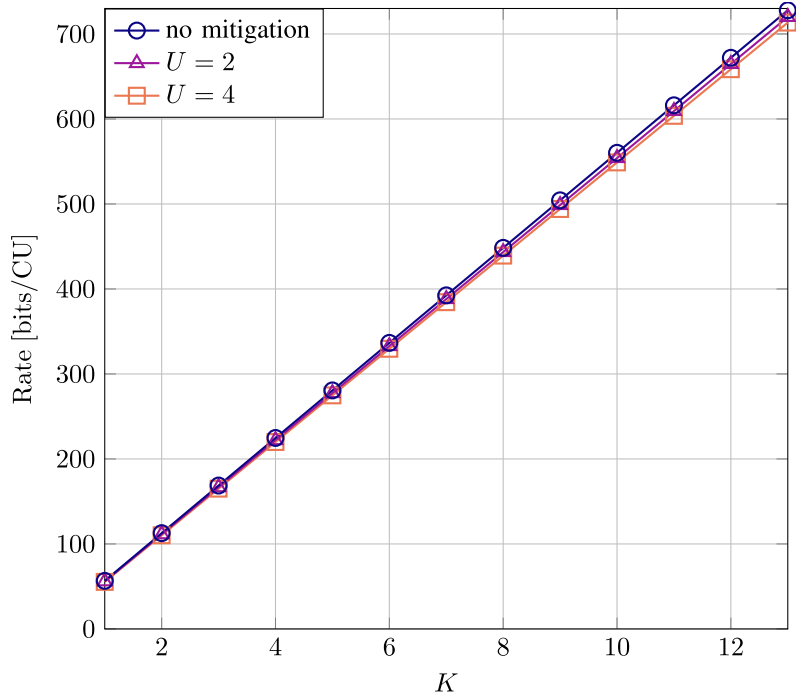}
\caption{Rate over number of subbands for $U=2$ and 4, and comparison with the rate achieved without the attack mitigation.}
\label{fig:mitigationRate}
\end{figure}
 
Lastly, we show the performance of our mitigation solution, wherein the UE selects one out of the $U$ precoders that provide the $U$ highest rates. \figurename~\ref{fig:mitigationRmse} shows the RMSE obtained with  $U=2$ and 4, with $N = 16$, $\bar{N} = 1$, codebook mode 2, and all other parameters as reported in Table~\ref{tab:numericalParameters}. Note that the conventional receiver without mitigation corresponds to $U=1$, where the chosen precoder maximizes the rate. We see that even with $U=2$ or $U=4$, the RMSE significantly increases, thus confirming that this technique is effective in protecting the location information privacy. 

Note also that, for a high number of subbands $K$, $U=4$ leads to a better localization accuracy than $U=2$. The reason is that with low $U$ not all feedback vectors are optimal at any location, and this leads to a more similar occurrence of the different precoders when $U$ increases. 

Now, the mitigation approach may yield a reduction of the downlink data rate, due to the use of a suboptimal precoder at the gNB. Therefore, \figurename~\ref{fig:mitigationRate} shows the rate (\ref{ratest}) achieved under the mitigation approach for $U = 2$, and 4, under the same setting of \figurename~\ref{fig:mitigationRmse}. Here, feedback mode 2 is used. For reference, we also include the case without mitigation. We note that the rate reduction is negligible. We can then conclude that the mitigation approach is very successful and can be adopted by receivers using the current standard and being recommended for all receivers in future releases of the standard. 
 
\section{Conclusions}\label{sec:Conclusion}

In this paper, we have proposed a novel attack, by which an attacker localizes a UE by eavesdropping the precoding feedback signal, which is transmitted in clear in the current 3GPP standard. We have detailed the operations to be performed by the attacker and analyzed the localization RMSE achieved using various feedback modes. We also proposed a mitigation technique, wherein the UE randomly selects one of the precoders among those providing the highest rates. Numerical results confirm that the attack is effective against the mandatory feedback mode~2 of the 3GPP standard, being able to localize the user within a few decimeters in a circular cell of radius~$\SI{25}{\meter}$. Moreover, asymptotic results provide an insight into the trends of the RMSE, as the number of subbands, antennas, or clusters increases. Lastly, the mitigation solution has proven to be very effective by increasing the localization RMSE  to multiple meters, with a negligible reduction  of the downlink data rate.  

\bibliographystyle{IEEEtran}
\bibliography{main.bib}
\end{document}